\documentclass[12pt]{article}
\usepackage{amsfonts,amssymb,epsfig}
\usepackage[latin1]{inputenc}
\usepackage[english]{babel}
\usepackage{hyperref}
\usepackage{color}
\newtheorem{thm}{Th\'eor\`eme}[section]
\newtheorem{cor}[thm]{Corollaire}
\newtheorem{lem}[thm]{Lemme}
\newtheorem{pro}[thm]{Proposition}
\newtheorem{dfn}[thm]{D\'efinition}
\newtheorem{rmk}[thm]{Remark}
\newtheorem{expl}[thm]{Exemple}

\def\dessous#1\sous#2{\mathrel{\mathop{\kern0pt#2}\limits_{#1}}}

\newcommand{\N}{\mathbb N}

\newcommand{\1}{1 \! \! {\rm I}}

\newcommand{\beq}{\begin{eqnarray}}
\newcommand{\eeq}{\end{eqnarray}}
\newcommand{\bpro}{\begin{pro}}
\newcommand{\epro}{\end{pro}}
\newcommand{\blem}{\begin{lem}}
\newcommand{\elem}{\end{lem}}
\newcommand{\bdfn}{\begin{dfn}}
\newcommand{\edfn}{\end{dfn}}
\newcommand{\bcor}{\begin{cor}}
\newcommand{\ecor}{\end{cor}}
\newcommand{\bthm}{\begin{thm}}
\newcommand{\ethm}{\end{thm}}
\newcommand{\bex}{\begin{expl}}
\newcommand{\eex}{\end{expl}}
\newcommand{\brmk}{\begin{rmk}}
\newcommand{\ermk}{\end{rmk}}
\newcommand{\benum}{\begin{enumerate}}
\newcommand{\eenum}{\end{enumerate}}
\newcommand{\bitem}{\begin{itemize}}
\newcommand{\eitem}{\end{itemize}}

\begin{document}
\begin{center}
{\Large\bf {Coherent states for a system of an electron moving in a  plane: case of discrete spectrum }}

 \vspace{0.5cm}

 Isiaka Aremua$^{1,2}$ and Laure Gouba$^{3}$ 

 $^{1}${\em
Universit\'{e} de Lom\'{e} (UL), Facult\'{e} Des Sciences  (FDS), D\'{e}partement de Physique} \\
{\em Laboratoire de Physique des Mat\'eriaux et des Composants \`a Semi-Conducteurs}\\
{\em Universit\'{e} de Lom\'{e} (UL), 01 B.P. 1515  Lom\'{e} 01, Togo.}\\  
$^{2}$International Chair of Mathematical Physics
 and Applications. \\
 {\em ICMPA-UNESCO Chair,  University of Abomey-Calavi}\\
 {\em 072 B.P. 50 Cotonou, Republic of Benin. E-mail: claudisak@gmail.com}\\
 
 $^{3}${\em
The Abdus Salam International Centre for Theoretical Physics (ICTP),
Strada Costiera 11, I-34151 Trieste Italy. E-mail: laure.gouba@gmail.com}

\vspace{1.0cm}

\today

\begin{abstract}
\noindent
In this work, we construct  different classes of  coherent states related to a quantum system,  recently studied in \cite{aremua-gouba},  of an electron moving in a  plane in uniform external magnetic and electric fields which possesses both discrete and continuous spectra. The eigenfunctions are realized as an orthonormal basis of a suitable Hilbert
space appropriate for building the related coherent states. These latter  are achieved in the context where we consider both spectra purely discrete obeying the criteria that a family of coherent states must satisfies.
\end{abstract}

\end{center}


\setcounter{footnote}{0}

\section{Introduction}

From a generalization of  the definition of canonical coherent states, Gazeau and Klauder proposed a method to construct temporally stable coherent states for a quantum system with one degree of
freedom \cite{gazeau-klauder}. Then, in the literature,  the method has been explored  for different kinds of  quantum systems with several degrees of
freedom. See for example \cite{antoine-gazeau-klauder, gazeau-novaes, thirulo} and references therein. Also, in some previous works, motivated by these developments, multidimensional vector coherent states  have been performed for Hamiltonians describing the nanoparticle dynamics in terms of a system of interacting bosons and fermions \cite{aremuannanovcs}; from a matrix (operator) formulation of the Landau problem and
the corresponding Hilbert space, an analysis of various  multi-matrix vector coherent states  extended to diagonal matrix
domains has been performed on the basis of Landau levels \cite{aremuaastp}. Besides, the motion of an electron in a noncommutative
$(x,y)$ plane, in a constant magnetic field background coupled with a harmonic potential has been 
examined with the relevant vector coherent states constructed and discussed \cite{aremuajnmp}. 

Following the method developed in  \cite{gazeau-klauder, gazeau-novaes}, we investigate in a recent work \cite{aremua-gouba} by considering Landau levels, various classes of   coherent states as in \cite{ab1, thirulo, gouba} arising
from  physical Hamiltonian describing a charged particle in an electromagnetic field, by introducing  additional parameters useful for handling  discrete and continuous spectra of the Hamiltonian.
In this work, we consider 
Consider an electron moving in a  plane $(x,y)$ in the uniform external
electric field  $\overrightarrow{E} =-\overrightarrow{\nabla}\Phi(x,y)$ and the uniform external
magnetic field $\overrightarrow{B}$ which is perpendicular to the
plane  described by the Hamiltonian \cite{aremua-gouba}
\beq{\label{es0}}
H = \frac{1}{2m}\left(\overrightarrow{p} +
\frac{e}{c}\overrightarrow{A}\right)^{2} - e\Phi.
\eeq
We briefly recall here a summary of results where the details are given in \cite{aremua-gouba}.
In the symmetric gauge $\overrightarrow{A} = \left(\frac{B}{2}y, -\frac{B}{2}x \right)$ with  the scalar potentiel given by $ \Phi(x,y) = -Ey$, 
the corresponding  classical Hamiltonian, obtained from (\ref{es0}), denoted by $
H_{1}$, reads 
\beq{\label{es1}} 
H_{1}(x,y,p_x,p_y)   =
\frac{1}{2m}\left[\left(p_{x} + \frac{eB}{2c}y\right)^{2} +
\left(p_{y} - \frac{eB}{2c}x\right)^{2} \right] + eEy. 
\eeq
The Hamiltonian $\hat H_1$ can be then re-expressed as follows:
\beq\label{es7}
\hat H_1 = \frac{1}{4m}\left(b^{\dag}b + bb^{\dag}\right) - \frac{\lambda}{2m}\left(d^{\dag} + d\right) - \frac{\lambda^{2}}{2m}
\eeq
and splits into two commuting parts in the following manner:
\beq 
\hat H_{1} = \hat H_{1_{OSC}} -  \hat T_1, 
\eeq
where  $\hat H_{1_{OSC}}$ denotes the harmonic oscillator part 
\beq
\hat H_{1_{OSC}} = \frac{1}{4m}(b^{\dag}b + bb^{\dag}),
\eeq 
while the
part linear in $d$ and  $d^{\dag}$ is given by 
\beq
\label{tfunc1}
\hat T_1 =
\frac{\lambda}{2m}(d^{\dag} + d) + \frac{\lambda^{2}}{2m}. 
\eeq 
Therefore,
the eigenvectors and the energy spectrum of
the Hamiltonian $\hat H_{1}$ are determined by the following formulas:
\beq{\label{es17}} 
\Psi_{n, \alpha} &=& \Phi_{n} \otimes
\phi_{\alpha} \equiv |n,\alpha\rangle, \cr \cr \mathcal
E_{(n,\alpha)} &=& \frac{\hbar \omega_{c}}{2}(2n + 1) - \frac{\hbar
\lambda}{m}\alpha - \frac{\lambda^{2}}{2m},  \;\;\; \;\; n= 0, 1, 2,
\dots. 
\eeq
In the symmetric gauge $\overrightarrow{A} = \left(-\frac{B}{2}y,\frac{B}{2}x \right)$ with the scalar potential  given by\- $\Phi(x,y) = -E x$,
the classical Hamiltonian $H$ in equation  (\ref{es0}) becomes
\beq{\label{ei19}}
H_{2}(x,y,p_x,p_y)  =
\frac{1}{2m}\left[\left(p_{x} - \frac{eB}{2c}y\right)^{2} +
\left(p_{y} + \frac{eB}{2c}x\right)^{2}\right] + eEx.
\eeq 
The Hamiltonian operator $\hat H_{2}$  can be then written as
\beq{\label{ei24}} 
\hat H_{2} =
\frac{1}{4m}(\mathfrak b^{\dag}\mathfrak b + \mathfrak b\mathfrak b^{\dag}) - \frac{\lambda}{2m}(\mathfrak d^{\dag} +
\mathfrak d) - \frac{\lambda^{2}}{2m},
\eeq 
with the harmonic oscillator part is given by 
\beq
\hat H_{2_{OSC}} = \frac{1}{4m}(\mathfrak b^{\dag}\mathfrak b + \mathfrak b\mathfrak b^{\dag})
\eeq
and the linear part by 
\beq
\hat T_{2} = \frac{\lambda}{2m}(\mathfrak d^{\dag} +
\mathfrak d) + \frac{\lambda^{2}}{2m}.
\eeq
The eigenvectors and the eigenvalues of the
Hamiltonian $\hat H_{2}$, as previously determined for $\hat H_{1}$, are obtained  as 
\beq{\label{eig003}} 
\Psi_{l, \alpha} &=& \Phi_{l} \otimes
\phi_{\alpha} \equiv |l,\alpha\rangle, \cr \cr \mathcal
E_{(l,\alpha)} &=& \frac{\hbar \omega_{c}}{2}(2l + 1) - \frac{\hbar
\lambda}{m}\alpha - \frac{\lambda^{2}}{2m} \;\;\; \;\; l= 0, 1, 2,
\dots. 
\eeq

The eigenvectors denoted $|\Psi_{nl}\rangle := |n, l\rangle = |n\rangle \otimes |l\rangle$ of $\hat H_{1_{OSC}} $ can be so chosen that they are also
the eigenvectors of $\hat H_{2_{OSC}}$, since $[\hat H_{1_{OSC}}, \hat H_{2_{OSC}}] = 0, $  as follows:
\beq{\label{equa37}}
\hat H_{1_{OSC}}|\Psi_{nl}\rangle  = \hbar \omega_{c} \left(n + \frac{1}{2}\right)
|\Psi_{nl}\rangle, \,\,  \hat H_{2_{OSC}}|\Psi_{nl}\rangle  = \hbar
\omega_{c} \left(l + \frac{1}{2}\right) |\Psi_{nl}\rangle, \; n, l  = 0,1,2,\dots
\eeq
so that $\hat H_{2_{OSC}}$ lifts the degeneracy of $\hat H_{1_{OSC}}$ and vice versa.

The present paper is a direct continuation of our work in reference \cite{aremua-gouba}, where we construct different classes of coherent states corresponding to the case of discrete spectrum.

The  paper is organized as follows. The section \ref{sec2} is
devoted to the construction of coherent states  for the quantum 
Hamiltonian possessing  purely  discrete spectrum by following the { method} developped in \cite{gazeau-klauder, gazeau-novaes}. The section \ref{sec3} is about the coherent states of the unshifted Hamiltonians $H_1$  and $H_2$ defined through multiple summations. In section \ref{sec4} we construct coherent states  related to the Hamiltonian $H_{1_{OSC}} - H_{2_{OSC}}$. An outlook is given is section \ref{sec5}.

\section{Coherent states for shifted Hamiltonians with more than one degree of freedom}\label{sec2}

In this section, we construct various classes of coherent states  for   Hamiltonian operators that admit discrete eigenvalues and eigenfunctions in  appropriate separable Hilbert spaces as elaborated in \cite{ab1, gazeau-novaes, thirulo}.

Let  $\mathfrak{H}_D$ be spanned by the eigenvectors $|\Psi_{nl}\rangle \equiv |n,l\rangle$ of $H_{1_{OSC}}$ and $H_{2_{OSC}}$ provided by (\ref{equa37}). Let us consider  
 $I_{\mathfrak H^l_D}, I_{\mathfrak H^n_D}$ the identity operators on the subspaces $\mathfrak H^n_D,\: \mathfrak H^l_D$ of $\mathfrak{H}_D$ such that 
\beq\label{identsubs}
\sum_{n=0}^{\infty}|\Psi_{nl}\rangle \langle \Psi_{nl}| = I_{\mathfrak H^l_D}, \quad \sum_{l=0}^{\infty}|\Psi_{nl}\rangle \langle \Psi_{nl}| = I_{\mathfrak H^n_D}.
\eeq
Since the Hamiltonians considered are formed by self adjoint operators that act in infinite dimensional Hilbert spaces, from 
the equations (\ref{es17}) and (\ref{eig003}), we set
\beq
\mathcal
E_{n,\alpha_k} &=& \frac{\hbar \omega_{c}}{2}(2n + 1) - \frac{\hbar
\lambda}{m}\alpha_k - \frac{\lambda^{2}}{2m}, \;\;\; \;\; n,k = 0, 1, 2,
\dots  \cr
\mathcal
E_{l,\alpha_k} &=& \frac{\hbar \omega_{c}}{2}(2l + 1) - \frac{\hbar
\lambda}{m}\alpha_k - \frac{\lambda^{2}}{2m}, \;\;\; \;\; l,k = 0, 1, 2,
\dots
\eeq
that define families of discrete eigenvalues associated with the eigenvectors $\{|\Psi_{nl}\rangle \otimes |\alpha_{k}\rangle \}_{n,l,k=0}^{\infty} $, forming an orthonormal basis of the separable Hilbert space $\hat{\mathfrak H} = \mathfrak H_{D} \otimes \tilde{\mathfrak H}$, with $\tilde{\mathfrak H}$ spanned by the states $\{|\alpha_k\rangle\}_{k=0}^{\infty}$.
\subsection{Coherent states of the shifted Hamiltonians
}
The eigenenergy of these shifted Hamiltonians can be written as
\beq{\label{el14}}
\mathcal E'_{n,\alpha_{k}} = \mathcal E_{n_{OSC}} -  \mathcal E_{\alpha_{k}} -
\left(\frac{\hbar \omega_{c}}{2} - \frac{\lambda^{2}}{2m} \right) = \hbar \omega_{c} n - \frac{\hbar \lambda }{m} \alpha_{k}.
\eeq
Then, the condition $\mathcal E'_{n,\alpha_{k}} \geq 0$,  for all $(k,n) \in \N \times \N^{*}$, requires
\beq{\label{cond}}
0 \leq \alpha_{k} \leq \frac{m \omega_{c}}{\lambda}.
\eeq
Let us suppose $\alpha_{k}$ fixed, then set
\beq{\label{val001}}
\mathcal E'_{n,\alpha_{k}} = \hbar \omega_{c} \left(n - \frac{\lambda}{m\omega_{c}}\alpha_{k} \right), \,\,\,
\kappa = \hbar \omega_{c}.
\eeq
We define
\beq
\rho(n) := \mathcal E'_{1,\alpha_{k}} \mathcal E'_{2,\alpha_{k}} \cdots \mathcal E'_{n,\alpha_{k}},
\eeq
with the increasing order $\mathcal E'_{1,\alpha_{k}} < \mathcal E'_{2,\alpha_{k}} < \cdots <
\mathcal E'_{n,\alpha_{k}}$,
such that
\beq{\label{val02}}
\rho(n) = \prod_{q=1}^{n} \hbar \omega_{c} \left(q - \frac{\lambda}{m\omega_{c}}\alpha_{k}\right) =
\kappa^{n}(\gamma)_{n}, \,\,\, \gamma = 1 - \frac{\lambda}{m\omega_{c}}\alpha_{k},
\eeq
where $(\gamma)_{n}$ is the Pochhammer symbol, with $(\gamma)_{n} = \gamma (\gamma + 1)(\gamma + 2)
\dots (\gamma +n -1)$.

Suppose now $n$ fixed and set
\beq
\mathcal E'_{n,\alpha_{k}} = \frac{\hbar \lambda}{m} \left(\frac{m\omega_{c}}{\lambda} n  - \alpha_{k}\right),
\,\,\,
\xi = \frac{\hbar \lambda}{m}
\eeq
allowing the definition of the quantity
\beq
\rho(\alpha_{k}) := \mathcal E'_{n,\alpha_{1}} \mathcal E'_{n,\alpha_{2}} \cdots \mathcal E'_{n,\alpha_{k}},
\eeq
with $\mathcal E'_{n,\alpha_{1}} < \mathcal E'_{n,\alpha_{2}} < \cdots <
\mathcal E'_{n,\alpha_{k}}$ and 
 $\epsilon_{k} ! = \epsilon_{k}\epsilon_{k-1}\cdots \epsilon_{1}$, such that
\beq\label{barrho}
\bar \rho(k) =
\prod_{q=1}^{k}\frac{\hbar \lambda}{m} \left( \frac{m\omega_{c}}{\lambda} n  - \alpha_{q}\right)
= \epsilon_{k} !\, \xi^{k}, \,\,\, \epsilon_{1}:=\frac{m\omega_{c}}{\lambda} n  - \alpha_{1}.
\eeq
\subsubsection{Coherent states with one degree of freedom}
Let  $l$ and $n$ fixed. Define the coherent states   for the Hamiltonian $ H_{1_{osc}} - T_{1} -  \left(\frac{\hbar \omega_{c}}{2} - \frac{\lambda^{2}}{2m}\right) I_{\hat{\mathfrak H}}$, where $\alpha_{k}$ satisfies (\ref{cond}), with one degree of freedom. Denoting them with the fixed index $l$, they are given from (\ref{barrho}), with $K \geq 0$ and $0 \leq \delta < 2\pi$, by
\beq\label{csoned000}
|K,\delta;l\rangle = \mathcal N(K,\delta;l)^{-1/2}\sum_{k=0}^{\infty}\frac{K^{k/2}
e^{-i \mathcal E'_{n,\alpha_{k}}\delta}}{\sqrt{\bar \rho(k)} }
|\Psi_{nl}\rangle \otimes |\epsilon_{k}\rangle.
\eeq
With the normalization condition
\beq
\langle K,\delta;l|K,\delta;l\rangle = 1,
\eeq
the normalization constant is determined such that we must have
\beq
\langle K,\delta;l|K,\delta;l \rangle = \mathcal N(K,\delta;l)^{-1}\sum_{k=0}^{\infty}\frac{K^{k}}{\epsilon_{k} !\xi^{k}} < \infty.
\eeq
Thus if $\lim_{k \to \infty}\epsilon_{k} = \epsilon$, we need to restrict $K$ to $0 \leq K < L = \sqrt{\epsilon}$ for the convergence of the above series. In this case we have
\beq{\label{eq2}}
\mathcal N(K,\delta;l) = \sum_{k=0}^{\infty}\frac{K^{k}}{\epsilon_{k} !\xi^{k}}.
\eeq
\bpro\label{prop4.2}
For fixed $l$ and $n$, let us write the following measures
\beq{\label{mes4}}
d\mu(K,\delta) = d\nu(K)d\mu(\delta) &=& \mathcal N(K,\delta;n)\varpi(K)dK\frac{d\delta}{2\pi}.
\eeq
Then,
on the Hilbert subspace $\mathfrak H^{nl}_{D}$ of  ${\mathfrak  H_{D}}$, the coherent states satisfy the  resolution of the identity
given by
\beq{\label{el25}}
\int_{0}^{L}\int_{0}^{2\pi}|K,\delta;l\rangle  \langle K,\delta;l| \; d\mu(K,\delta)
= I_{\mathfrak H^{nl}_{D}} \otimes I_{\tilde{\mathfrak H}}.
\eeq
\epro
{\bf Proof.} See in the Appendix.
$\hfill{\square}$
\subsubsection{Coherent states with two degrees of freedom}
Let  $\alpha_{k}$ fixed. We obtain infinite component vector coherent states \cite{ab1}, with two degrees of freedom, each component counting the infinite degeneracy of the energy level of the harmonic oscillator shifted eigenvalues $\mathcal E'_{n}
= \hbar \omega_{c} n $.
Taking $J \geq 0, \,  J' \geq 0$ and $0 \leq \theta, \theta' < 2\pi $,
the coherent states are given by
\beq{\label{el22}}
|J,\theta;J',\theta';l\rangle
&=& \mathcal N(J;\alpha_{k})^{-1/2}
\mathcal N(J';\alpha_{k})^{-1/2} J'^{l/2} e^{i \mathcal E'_{l,\alpha_{k}}\theta'} \cr
&&\times\sum_{n=0}^{\infty}\frac{J^{n/2}e^{-i \mathcal E'_{n,\alpha_{k}}\theta}}{\sqrt{\rho(n)\rho(l)}}
|\Psi_{nl}\rangle \otimes |\alpha_{k}\rangle,
\eeq
where 
\beq 
\mathcal N(J,\alpha_{k}) = \sum_{n=0}^{\infty}\frac{J^{n}}{(\gamma)_{n} \kappa^{n}} =  \,_{1}F_{1}\left(1;\gamma;\frac{J}{\kappa} \right).
\eeq
\bpro\label{prop4.1}
Provided the measures
\beq{\label{mes3}}
d\eta(J,\theta) = d\rho(J)d\mu(\theta) &=& \frac{1}{\kappa^{\gamma}\Gamma(\gamma)}   \, _{1}F_{1}\left(1;\gamma;\frac{J}{\kappa}\right)
e^{-J/\kappa}J^{\gamma - 1}dJ\frac{d\theta}{2\pi}
\eeq
with fixed $\alpha_{k}$,  the resolution of the identity 
 can be expressed as follows:
\beq{\label{el24}}
\int_{0}^{\infty}\int_{0}^{\infty}
\int_{0}^{2\pi}\int_{0}^{2\pi}|J,\theta;J',
\theta';l\rangle \langle J,\theta;J',\theta';l| d\eta(J,\theta) d\eta(J',\theta') &=& I_{\mathfrak H^l_D} \otimes
 I_{\tilde{\mathfrak H}_{k}},
\eeq
where $ I_{\mathfrak H^l_D}$  is defined previously in (\ref{identsubs}) 
and
$\tilde{\mathfrak H}_{k}$ the subspace  of $\tilde{\mathfrak H}$, ${\tilde{\mathfrak H}}$ being  spanned by $\{|\alpha_k\rangle\}_{k=0}^{\infty}$ with
\beq
\sum_{k=0}^{\infty}|\alpha_k\rangle \langle \alpha_k| = \sum_{k=0}^{\infty}I_{\tilde{\mathfrak H}_{k}} = I_{\tilde{\mathfrak H}}.
\eeq
\epro
{\bf Proof.} See in the Appendix.$\hfill{\square}$

\bpro\label{prop4.3}
The coherent states defined  in (\ref{el22}) and (\ref{csoned000}) satisfy the temporal stability property given
as follows:
\beq
e^{-i H'_{1}t}|J,\theta;J',\theta';l\rangle &=&
|J,\theta + t;J',\theta';l\rangle,
\cr
e^{-i H'_{1}t}|K,\delta;l\rangle &=& |K,\delta +t;l\rangle,
\eeq
with $H'_{1} = H_{1_{osc}} - T_{1} -  \left(\frac{\hbar \omega_{c}}{2} - \frac{\lambda^{2}}{2m}\right) I_{\hat{\mathfrak H}}$.
\epro
{\bf Proof. } See in the Appendix.
$\hfill{\square}$
\begin{rmk}{\label{rem1}}
Note that since in the equation (\ref{el14}), we have $\mathcal E'_{0,\alpha_{k}} \neq 0$, the coherent states
(\ref{csoned000}) and (\ref{el22}) 
cannot satisfy the action identity. In this case, we phrase the resulting coherent states as ''temporally stable coherent states``.
\end{rmk}

\section{Coherent states for  unshifted Hamiltonians $H_{1}$ and $H_{2}$ defined through multiple summations}\label{sec3}

In this paragraph, two types of temporally stable coherent states are constructed in line with the general scheme developed in \cite{gazeau-novaes}, (see also \cite{thirulo}). The first type is defined as a tensor product of two classes of coherent states   with one and two degrees of freedom by setting  $\rho_{1}(n), \mathcal E_{n}, \mathcal E_{\alpha_{k}}$ and $\bar \rho(k)$ as independent  quantities. The second type, which cannot be considered as a tensor product of vectors with one and two degrees of freedom, is defined by letting that one sum depends on the other through the same quantities.
\bitem
\item [1-] When the summations are independent

Let us set $\mathcal E_{n,\alpha_{k}} = \mathcal E_{n} + \mathcal E_{\alpha_{k}}$, where
\beq
\mathcal E_{n} = \hbar \omega_{c}n, \qquad  \mathcal E_{\alpha_{k}} = -\frac{\hbar \lambda}{m}\alpha_{k} := \mathcal E'_{k}.
\eeq
Then, the conditions of positivity required for the eigenvalues  $\mathcal E'_{k}$  imposes  $\alpha_{k} \leq 0$.
 Setting $\rho_{{1}}(n) = \mathcal E_{1} \mathcal E_{2}
\cdots \mathcal E_{n}$, $\bar \rho(k) = \mathcal E'_{1}\mathcal E'_{2} \cdots \mathcal E'_{k}$
and  $\epsilon_{k} = \rho_{2}(k)/\rho_{2}(k-1)$, where $\epsilon_{k} := -\alpha_{k}$, for $k = 1, 2, 3,\dots$
 leads to $\rho_{2}(k) = \epsilon_{k}\epsilon_{k-1}\cdots \epsilon_{1} = \epsilon_{k} ! $
 with, by convention, $\epsilon_{0} ! = 1$.
Given the relations $\mathcal E_{1} < \mathcal E_{2}< \cdots
< \mathcal E_{n}$ and $\mathcal E'_{1} < \mathcal E'_{2} < \cdots \mathcal E'_{k}$, we can rewrite
\beq{\label{es21}}
\rho_{1}(n) &=& \prod_{k=1}^{n}\hbar \omega_{c} k = n ! \,\kappa^n, \; \kappa = \hbar \omega_{c},
\cr
\bar \rho(k)&=& \prod_{q=1}^{k}\frac{\hbar \lambda}{m}\epsilon_{q}  =
\upsilon^{k}\rho_{2}(k) = \epsilon_{k} !\upsilon^{k}, \quad \upsilon = \frac{\hbar \lambda}{m}.
\eeq
Under these considerations,  the  coherent states  for the Hamiltonians $H_{1_{OSC}} - T_{1}$
and $H_{2_{OSC}} - T_{2}$, when taking into account the degeneracies of the Landau levels as before,
are defined with three degrees of freedom as a tensor product of two coherent states defined  with one and two
degrees of freedom, respectively, as follows:
\beq{\label{es25}}
|J,\theta;J',\theta';K,\delta;l\rangle
&=& [\mathcal N(J)\mathcal N(J')]^{-1/2}\mathcal N(K)^{-1/2}J'^{l/2}e^{i \mathcal E_{l}\theta'}\sum_{n=0}^{\infty}\frac{J^{n/2}e^{-i \mathcal E_{n}\theta}}{\sqrt{\rho_{1}(l)\rho_{1}(n)}}
\cr
&&\times\sum_{k=0}^{\infty}\frac{K^{k/2}e^{-i \mathcal E'_{k}\delta}}{\sqrt{\bar \rho(k)}}|\Psi_{nl}\rangle \otimes |\epsilon_{k}\rangle,
\crcr
|J,\theta;J',\theta';K,\delta;n\rangle
&=& [\mathcal N(J)\mathcal N(J')]^{-1/2}\mathcal N(K)^{-1/2} J^{n/2}e^{-i \mathcal E_{n}\theta}\sum_{l=0}^{\infty}
\frac{J'^{l/2}e^{i \mathcal E_{l}\theta'}}{\sqrt{\rho_{1}(n)\rho_{1}(l)}} \cr
&&\times \sum_{k=0}^{\infty}\frac{K^{k/2}e^{-i \mathcal E'_{k}\delta}}{\sqrt{\bar \rho(k)}}|\Psi_{nl}\rangle \otimes |\epsilon_{k}\rangle.
\eeq
\bpro
The normalization requirements
\beq{\label{es23}}
\sum_{l=0}^{\infty}\langle J,\theta;J',\theta';K,\delta;l|J,\theta;J',\theta';K,\delta;l \rangle = 1
\eeq	
are such that we must have the relation (\ref{eq2}). Then, we obtain
\beq{\label{es24}}
 \mathcal N(J)= \sum_{n=0}^{\infty}\frac{J^{n}}{\rho_{1}(n)} =  e^{J/\kappa},
\quad 
\mathcal N(K) = \sum_{k=0}^{\infty}\frac{K^{k}}{\epsilon_{k} !\xi^{k}}.
\eeq
\epro
\bpro\label{prop4.6}
The coherent states (\ref{es25}), 
satisfy, on the Hilbert space $\mathfrak H_{D} \otimes \tilde{\mathfrak H}$,
the following resolutions of the identity:
\beq{\label{es28}}
&&\int_{0}^{\infty}\int_{0}^{\infty}\int_{0}^{L}
\int_{0}^{2\pi}\int_{0}^{2\pi}\int_{0}^{2\pi}|J,\theta;J',
\theta';K,\delta;l\rangle \langle J,\theta;J',\theta';K,\delta;l| d\eta(J,\theta) \cr &&\times d\eta(J',\theta')d\mu(K,\delta)= \1_{\mathfrak H^{l}_{D}} \otimes \1_{\tilde{\mathfrak H}}.
\eeq
\epro
{\bf Proof.} See in the Appendix.
$\hfill{\square}$
\bpro
The coherent states (\ref{es25}) 
are temporally stable, i.e.,
\beq
e^{-i (H_{1_{OSC}} - T_{1})t}|J,\theta;J',\theta';K,\delta;l\rangle = |J,\theta + t;J',\theta';K,\delta + t;l\rangle.
\eeq
\epro
{\bf Proof.} See that of Proposition
4.2

$\hfill{\square}$
\item [2-] When the summations depend one on the other

For fixed $\alpha_{k}$, with $\mathcal E'_{n,\alpha_{k}}$ given in (\ref{val001}), let us set
\beq
\rho(n,\alpha_{k}) = \mathcal E'_{n,\alpha_{1}}\mathcal E'_{n,\alpha_{2}}\dots \mathcal E'_{n,\alpha_{k}}.
\eeq
From (\ref{val02}), one has
\beq
\rho(n,\alpha_{k}) = \kappa^{n}(\gamma)_{n}.
\eeq
The definition (\ref{es21}) gives
\beq
\rho_{1}(l) = l !\kappa^{n}.
\eeq
Under the condition  $\alpha_{k} \leq 0$, one can define the following coherent states:
\beq{\label{es27}}
|J,\theta;J',\theta';K,\delta;l\rangle
&=& \mathcal N(J')^{-1/2}\mathcal N(J,\alpha_{k})^{-1/2}J'^{l/2}e^{i \mathcal E'_{l,\alpha_{k}}\theta'}\sum_{n=0}^{\infty}\frac{J^{n/2}e^{-i \mathcal E'_{n,\alpha_{k}}\theta}}{\sqrt{\rho_{1}(l)\rho(n,\alpha_{k})}}
\cr
&& \times \mathcal N(K,J)^{-1/2}\sum_{k=0}^{\infty}\frac{K^{k/2}e^{-i \mathcal E'_{k}\delta}}{\sqrt{\bar \rho(k)}}|\Psi_{nl}\rangle \otimes |\epsilon_{k}\rangle.
\eeq
In order to obtain the normalization constant, let us compute the norm of the coherent states $|J,\theta;J',\theta';K,\delta;l\rangle$, with the conditions given in (\ref{es23}) leading to
\beq
&&\langle J,\theta;J',\theta';K,\delta;l|J,\theta;J',\theta';K,\delta;l\rangle \cr
&&=\mathcal N(J)^{-1}\sum_{l=0}^{\infty}\frac{J^{l}}{\rho_{1}(l)}\mathcal N(K,J)^{-1}\sum_{k=0}^{\infty}\frac{K^{k}}{\bar \rho(k)}\mathcal N(J,\alpha_{k})^{-1}\sum_{n=0}^{\infty}\frac{J^{n}}{\rho_{1}(n,\alpha_{k})} = 1 \nonumber
\\
\eeq
if, as done in (\ref{eq2}) and (\ref{es24}),
\beq
\mathcal N(J,\alpha_{k}) = \,_{1}F_{1}\left(1;\gamma;\frac{J}{\kappa} \right) \geq 1, \qquad \mathcal N(J') = e^{-J'/\kappa} \geq 1,
\eeq
and
\beq
\mathcal N(K,J) &=& \sum_{n=0}^{\infty}\frac{K^{k}}{\bar \rho(k)\mathcal N(J')\mathcal N(J,\alpha_{k})}.
\eeq
Then, we get
\beq
\mathcal N(K,J)
&=&  \sum_{k=0}^{\infty}\frac{K^{k}}{\epsilon_{k} !\xi^{k} \, e^{-J'/\kappa}\, _{1}F_{1}\left(1;\gamma;\frac{J}{\kappa}\right)
} \leq \sum_{k=0}^{\infty}
\frac{K^{k}}{\epsilon_{k} !\xi^{k}},
\eeq
which converges for all $ 0 \leq K \leq L = \sqrt{\epsilon}$.
\bpro\label{prop4.8}
The coherent states  (\ref{es27}) satisfy, on $\mathfrak H_{D} \otimes \tilde{\mathfrak H}$, the resolutions of the identity  given by
\beq
&&
\int_{0}^{\infty}\int_{0}^{\infty}\int_{0}^{L}
\int_{0}^{2\pi}\int_{0}^{2\pi}\int_{0}^{2\pi}|J,\theta;J',
\theta';K,\delta;l\rangle \langle J,\theta;J',\theta';K,\delta;l| d\eta(J,\theta) \cr \cr &&\times d\eta(J',\theta')d\mu(K,n;\delta)=  \1_{\mathfrak H^{l}_{D}} \otimes \1_{\tilde{\mathfrak H}}.
\eeq
\epro
{\bf Proof.} See in the Appendix. 
$\hfill{\square}$

Note that, as previously mentioned, the coherent states (\ref{es27}) are temporally stable.
\eitem

\section{Coherent states related to  the Hamiltonian $H_{1_{OSC}} - H_{2_{OSC}}$}\label{sec4}

\bitem
\item When $\alpha_{k}$ is fixed: the coherent states are defined on $\mathfrak H_{D} \otimes \tilde{\mathfrak H}$, in  an analogous way, by using the `bi-coherent states' (BCS) \cite{ab1} as follows:
\beq{\label{el19}}
|J,\theta;J',\theta'\rangle  &=& |J,\theta;J',\theta'\rangle ^{BCS} \otimes |\alpha_{k}\rangle \cr
                                    &=& \frac{1}{\left[\mathcal N(J) \mathcal N(J')
\right]^{1/2}}\sum_{n,l=0}^{\infty}\frac{J^{n/2}J'^{l/2}e^{-i (n \theta - l \theta')}}{\sqrt{n !l !}}
|\Psi_{nl}\rangle \otimes |\alpha_{k}\rangle
\eeq
or, by using the complex labels, as
\beq{\label{el23}}
|z,\bar{z'}\rangle  &=& |z,\bar{z'}\rangle^{BCS} \otimes |\alpha_{k}\rangle \cr
                     &=& e^{-\frac{|z|^{2} + |z'|^{2}}{2}}\sum_{n,l=0}^{\infty}
\frac{z^{n}\bar{z'}^{l}}{\sqrt{n !l !}}|\Psi_{nl}\rangle \otimes |\alpha_{k}\rangle.
\eeq

They
correspond to the multidimensional coherent states \cite{gazeau-novaes} of the Hamiltonian $H_{1_{OSC}} - H_{2_{OSC}}$.

The normalization condition is given by
\beq
\langle J,\theta;J',\theta'|J,\theta;J',\theta'\rangle = 1.
\eeq
\bpro
They satisfy, on the separable Hilbert space $\tilde{\mathfrak H}_{k}$,  the following resolution of the identity:
\beq
&&\int^{\infty}_{0}\int^{\infty}_{0}
\int_{0}^{2\pi}\int_{0}^{2\pi}|J,\theta;J',\theta'\rangle \langle J,\theta;J',\theta'|
 d\mu(\theta)d\mu(\theta') \cr
&& \times \mathcal N(J)\mathcal N(J')d\nu(J)d\nu(J') = \1_{\tilde{\mathfrak H}_{k}}.
\eeq

\epro
\eitem
\bitem
\item  When $\alpha_{k}$ is not fixed: the coherent states are denoted by $|J,\theta;J',\theta';\alpha_{k}\rangle $ or $|z,\bar{z'};\alpha_{k}\rangle $ and given by the same equations (\ref{el19}) and (\ref{el23}).

Here, the normalization condition is given by
\beq
\sum_{k=0}^{\infty}\langle J,\theta;J',\theta';\alpha_{k}|J,\theta;J',\theta';\alpha_{k} \rangle = 1.
\eeq
\bpro\label{prop4.10}
They satisfy, on the separable Hilbert space $\hat{\mathfrak H} = \mathfrak H_{D} \otimes \tilde{\mathfrak H}$,  the following resolution of the identity:
\beq
&&
\int^{\infty}_{0}\int^{\infty}_{0}
\int_{0}^{2\pi}\int_{0}^{2\pi}|J,\theta;J',\theta';\alpha_{k}\rangle \langle J,\theta;J',\theta';\alpha_{k}|
 d\mu(\theta)d\mu(\theta') \cr
&& \times\mathcal N(J)\mathcal N(J')d\nu(J)d\nu(J') =
\1_{\mathfrak H_{D}} \otimes \1_{\tilde{\mathfrak H}_k}.
\eeq
\epro
\eitem
{\bf Proof.} See in the Appendix.
$\hfill{\square}$
\bpro\label{prop4.11}
The coherent states (\ref{el19}) also satisfy the properties of temporal stability and action identity as stated in \cite{ab1}. In the situation of the coherent states $|J,\theta;J',\theta';\alpha_{k}\rangle$, these properties are given as below
\beq{\label{el20}}
e^{-i H t}|J,\theta;J',\theta';\alpha_{k}\rangle = |J,\theta + \omega_{c} t;J',\theta' + \omega_{c} t;\alpha_{k}\rangle
\eeq
\beq{\label{el21}}
\langle J,\theta;J',\theta';\alpha_{k}| H |J,\theta;J',\theta';\alpha_{k}\rangle = \omega_{c} (J-J')
\eeq
with $H = H_{1_{OSC}} - H_{2_{OSC}}$.
\epro
{\bf Proof.} See in the Appendix.
$\hfill{\square}$

\section{Outlook}\label{sec5}

The behaviour of an electron moving in a  plane in an  electromagnetic field background, arising in
the quantum Hall effect,  has been studied with the related Hamiltonian spectra  having  both discrete and continuous parts  provided. Also, an Hamiltonian in the case of an electric field depending simultaneously
on both x and y directions has  been discussed with his spectrum provided. The eigenfunctions have been  obtained as a countable set realizing an infinite dimensional appropriate Hilbert space. Various  coherent states have been  constructed  by considering shifted and unshifted spectra, respectively. Two kinds of coherent states classes have been obtained. The first kind,  with one degree of freedom,  is achieved by fixing each index counting the energy levels. The second kind is realized by taking  tensor product of
two classes of coherent states  with one and two degrees of freedom. 

The discussion can be extended for th case of the potential $V = E_{1}x + E_{2}y$.
Here, the uniform electric field is defined as ${\overrightarrow E} = (E_{1}, E_{2}, 0)$ with  the scalar potential $\Phi(x,y) = E_{1}x + E_{2}y = {\bf E}\cdot {\bf r}$,  
and the magnetic field given by ${\bf A} = \left(-\frac{B}{2}y, \frac{B}{2}x \right)$.
Then,  the Hamiltonian writes as
\beq{\label{ei19}}
H(x,y,p_x,p_y) = \frac{1}{2M}\left[\left(p_{x} - \frac{eB}{2c}y\right)^{2} + \left(p_{y} + \frac{eB}{2c}x\right)^{2}\right] - eE_{1}x - eE_{2}y.
\eeq
Let us  introduce the following pairs of  annihilation and creation operators  defined by
\beq{\label{es002}}
b^{\dag} = -2i P_{\bar{z}} + \frac{eB}{2c}Z - 2\lambda_{1}, \qquad 
b = 2i P_{z} +  \frac{eB}{2c}\bar{Z} - 2\lambda_{1},
\eeq
\beq{\label{es003}}
d = 2i P_{z}-\frac{eB}{2c}\bar Z, \qquad 
d^{\dag} = -2i P_{\bar z} - \frac{eB}{2c} Z,
\eeq
and
\beq{\label{ei20}}
\hat l = 2P_{z} - i \frac{eB}{2c}\bar{Z} + 2\lambda_{2}, \qquad
\hat l^{\dag}  =  2P_{\bar{z}} + i \frac{eB}{2c}Z + 2\lambda_{2},
\eeq
\beq{\label{ei21}}
\hat k^{\dag}  = 2P_{\bar{z}} - i \frac{eB}{2c}Z ,\qquad
\hat k = 2 P_{z} + i \frac{eB}{2c}\bar{Z},
\eeq
where $\lambda_{1} = \frac{McE_{1}}{B}$ and $\lambda_{2} = \frac{McE_{2}}{B}$. They satisfy the following commutation relations
\beq{\label{es4}}
[b, b^{\dag}] = 2 M \hbar \omega_{c}, \;\; [d^{^{\dag}}, d] = 2 M \hbar \omega_{c}, \;\;\;
[b,d^{\dag}] = 0, \;\;[b^{\dag}, d] = 0, \;\; [b^{\dag}, d^{\dag}] = 0, \;\; [b,d] = 0, 
\eeq
\beq{\label{es4}}
[\hat l, \hat l^{\dag}] = 2 M \hbar \omega_{c}, \;\; [\hat k^{\dag}, \hat k] = 2 M \hbar \omega_{c}, \;\;
[\hat k,\hat l^{\dag}] = 0, \;\;[\hat k^{\dag}, \hat l] = 0, \;\;[\hat k^{\dag}, \hat l^{\dag}] = 0, \;\;[\hat k,\hat l] = 0, 
\eeq
and
\beq
[b^{\dag}, \hat k] = 0 = [b,\hat k^{\dag}], \quad [\hat l^{\dag}, d] = 0 = [\hat l, d^{\dag}].
\eeq
The operator Hamiltonian  $\hat H$ is delivered as follows:
\beq{\label{es007}}
\hat H = \frac{1}{4M}( l^{\dag}l+ll^{\dag})+ \frac{|\xi|^2}{M} + \frac{1}{2M}(\xi l^{\dag} + \bar \xi l)\ + \frac{\lambda_{1}}{2M}(d^{\dag} + d) + 
\frac{\lambda^{2}_{1}}{M} + \frac{\lambda_{2}}{2M}(\hat k^{\dag} + \hat k) - \frac{\lambda^{2}_{2}}{M}
\eeq
where we use  the following relations 
\beq
b^{\dag} = -i (\hat l^{\dag} + 2\bar{\xi}), \quad  b= i (\hat l + 2\xi), \quad \xi = i \lambda_{1} - \lambda_{2},\quad 
d = i \hat{k}, \quad d^{\dag} = -i \hat k^{\dag}.
\eeq 
The eigenvalue equation $\hat H \Psi  = E \Psi$, $\Psi(r,\theta) = \varphi(r)e^{i l \theta}$,  provides 
the radial equation
\beq{\label{part00}}
\left(\partial^{2}_{r} + \frac{1}{r}\partial_{r} - \frac{l^{2}}{r^{2}}\right)\varphi(r) - \left(2Br  + 2Cr^{2} - 2\mathcal E\right)\varphi(r) = 0
\eeq
where
\beq
B=-\frac{M\omega_{c}}{\hbar^{2}}f_{1}(\theta), \quad C = \left(\frac{M\omega_{c}}{2\sqrt{2}\hbar}\right)^{2}, \quad \mathcal E = \frac{M}{\hbar^{2}}\left(E - \frac{\hbar l}{2}\omega_{c}\right)
\eeq
with  $f_{1}(\theta) = \lambda_{1}\cos{\theta} + \lambda_{2}\sin{\theta}.$ 
 The corresponding radial  eigenfunctions are obtained in terms of Heun functions as follows
\beq\label{eigfunc000}
\varphi(r) = A_1 r^{\frac{l_1 +1}{2}}e^{l_2(r)}{\mbox{HeunB}}\left\{l_1, l_3, l_4, 0, l_5(r)\right\} + A_2 r^{\frac{-l_1 +1}{2}}e^{l_2(r)}{\mbox{HeunB}}\left\{-l_1, l_3, l_4, 0, l_5(r)\right\}
\eeq
with 
\beq 
l_1 = 2l, l_2(r) = \frac{r}{\sqrt{2}}\left(\sqrt{C}r + \frac{B}{\sqrt{C}}\right), l_3 = i B\left(\frac{2}{C^3}\right)^{\frac{1}{4}}, l_4 = -\frac{4\mathcal{E}C + B^2}{2\sqrt{2}C^{\frac{3}{2}}}, l_5(r) = i r (2C)^{\frac{1}{4}},
\eeq
where $A_1$ and $A_2$ are constants. Thus, the solutions of the Schr\"{o}dinger equation $\hat H\Psi = E\Psi$ are obtained as the product of (\ref{eigfunc000}) by $e^{il\theta}$.
\section{Acknowlegments}
I. Aremua
would like to gratefully thank Professor A. S. d'Almeida for some valuable discussions. 

\section{Appendix}
{\bf Proof of Proposition \ref{prop4.2}}  From the definition (\ref{csoned000}), we have
\beq
|K,\delta;l\rangle  \langle K,\delta;l|
= 
|\Psi_{nl}\rangle \langle \Psi_{nl}| \otimes \mathcal N(K,\delta;l)^{-1} \sum_{k,q=0}^{\infty}\frac{K^{\frac{k + q}{2}}
e^{i (\mathcal E'_{n,\alpha_{q}} -
\mathcal E'_{n,\alpha_{k}})\delta}}{\sqrt{ \bar \rho(k)\bar \rho(q)}} |\epsilon_{k}\rangle \langle \epsilon_{q}|,
\eeq
which allows to write
\beq
&&\int_{0}^{L}\int_{0}^{2\pi}|K,\delta;l\rangle  \langle K,\delta;l|  d\mu(K,\delta)\cr
&&=
|\Psi_{nl}\rangle \langle \Psi_{nl}| \otimes \int_{0}^{L}
\sum_{k,q = 0}^{\infty}\frac{K^{\frac{k+q}{2}}}{\sqrt{\bar \rho(k) \bar \rho(q)}}
\int_{0}^{2\pi}e^{i (\mathcal E'_{n,\alpha_{q}} -  \mathcal E'_{n,\alpha_{k}} )\delta}
\frac{d\delta}{2\pi}\varpi(K)dK|\epsilon_{k}\rangle \langle \epsilon_{q}| \cr
\cr
&& = |\Psi_{nl}\rangle \langle \Psi_{nl}| \otimes \sum_{k=0}^{\infty}\int_{0}^{L} \frac{K^{k}}{\epsilon_{k} !\xi^{k}} \varpi(K)dK |\epsilon_{k}\rangle \langle \epsilon_{k}|.
\eeq
Assuming that the density $\varpi(K)$ satisfies the relation
\beq{\label{equa46}}
\int_{0}^{L} K^{k}\varpi(K)dK = \epsilon_{k} !\xi^{k},
\eeq
we get
\beq
\int_{0}^{L}\int_{0}^{2\pi}|K,\delta;l\rangle  \langle K,\delta;l|  d\mu(K,\delta)
= |\Psi_{nl}\rangle \langle \Psi_{nl}| \otimes\sum_{k=0}^{\infty}
|\epsilon_{k}\rangle \langle \epsilon_{k}| =  \1_{\mathfrak H^{nl}_{D}} \otimes \1_{\tilde{\mathfrak H}}.
\eeq
$\hfill{\square}$

{\bf Proof of Proposition \ref{prop4.1}}
Using  the definition (\ref{el22}) leads to
\beq
&&\int_{0}^{\infty}\int_{0}^{\infty}\int_{0}^{2\pi}\int_{0}^{2\pi}
|J,\theta;J',\theta';l\rangle \langle J,\theta;J',\theta';l| d\eta(J,\theta) d\eta(J',\theta') \cr
\cr
&& = \int_{0}^{\infty}\int_{0}^{\infty} \frac{J'^{l}}{\rho(l)} \frac{e^{-J'/\kappa}}{\kappa^{\gamma} \Gamma(\gamma)} J'^{\gamma-1}
\sum_{n,q = 0}^{\infty}\frac{J^{\frac{n+q}{2}}}{\sqrt{\rho(n)\rho(q)}}
\frac{J^{n}}{\kappa^{\gamma} \Gamma(\gamma)
} J^{\gamma-1}e^{-J/\kappa}\cr
&& \times\int_{0}^{2\pi}\int_{0}^{2\pi}e^{-i (\mathcal E'_{n,\alpha_{k}} -  \mathcal E'_{q,\alpha_{k}} )\theta}
\frac{d\theta}{2\pi}\frac{d\theta'}{2\pi}dJdJ'
|\Psi_{nl}\rangle \langle \Psi_{nl}| \otimes |\alpha_{k}\rangle \langle \alpha_{k}|.
\eeq
The  following relations
\beq
\int_{0}^{\infty} J^{n + \gamma-1}e^{-J/\kappa}
 dJ = \kappa^{n+\gamma}\Gamma(n + \gamma), \quad
\int_{0}^{\infty} J'^{l + \gamma-1}e^{-J'/\kappa}
dJ' = \kappa^{l+\gamma}\Gamma(l + \gamma)
,
\eeq
are satisfied, where we have used the inverse Mellin transform \cite{ismail, erdelyi}
\beq{\label{equa45}}
\int_{0}^{\infty}e^{-au}u^{s-1}du
=a^{-s}\Gamma(s),
\eeq
with $s = n+\gamma$ and $a = \frac{1}{\kappa}$. Thereby 
\beq
&&\int_{0}^{\infty}\int_{0}^{\infty}
\int_{0}^{2\pi}\int_{0}^{2\pi}|J,\theta;J',
\theta';l\rangle \langle J,\theta;J',\theta';l| d\eta(J,\theta) d\eta(J',\theta') \cr
&&=\sum_{n=0}^{\infty} |\Psi_{nl}\rangle \langle \Psi_{nl}| \otimes
|\alpha_{k}\rangle \langle \alpha_{k}| = 
\1_{\mathfrak H^l_D} \otimes \1_{\tilde{\mathfrak H}_{k}}.
\eeq
$\hfill{\square}$

{\bf Proof of Proposition \ref{prop4.3}} From the definition (\ref{el22}),  it follows that
\beq 
e^{-i H'_{1}t}|J,\theta;J',\theta';l\rangle
&=& \mathcal N(J,\theta;\alpha_{k})^{-1/2}
\mathcal N(J',\theta';\alpha_{k})^{-1/2} J'^{l/2} e^{i \mathcal E'_{l,\alpha_{k}}\theta'} \cr
&&\times \sum_{n=0}^{\infty}\frac{J^{n/2}e^{-i \mathcal E'_{n,\alpha_{k}}\theta}}{\sqrt{\rho(n)\rho(l)}
e^{-i H'_{1}t}|\Psi_{nl}\rangle \otimes |\alpha_{k}\rangle} \cr
&=& \mathcal N(J,\theta;\alpha_{k})^{-1/2}
\mathcal N(J',\theta';\alpha_{k})^{-1/2} J'^{l/2} e^{i \mathcal E'_{l,\alpha_{k}}\theta'} \cr
&&\times \sum_{n=0}^{\infty}\frac{J^{n/2}e^{-i \mathcal E'_{n,\alpha_{k}}(\theta + t)}}{\sqrt{ \rho(n)\rho(l)}}
|\Psi_{nl}\rangle \otimes |\alpha_{k}\rangle\cr
&=& |J, \theta + t;J',\theta';l\rangle.
\eeq
$\hfill{\square}$

{\bf Proof of Proposition \ref{prop4.6}.} Starting from (\ref{es25}),  with the measures given  by
\beq
d\eta(J,\theta) = e^{-J/\kappa}\kappa^{n}e^{-J}dJ\frac{d\theta}{2\pi}, \quad d\eta(J',\theta) = e^{-J'/\kappa}\kappa^{n}e^{-J'}dJ'\frac{d\theta'}{2\pi}
\eeq
and (\ref{mes4}), we get
\beq
&&\int_{0}^{\infty}\int_{0}^{\infty}\int_{0}^{L}
\int_{0}^{2\pi}\int_{0}^{2\pi}\int_{0}^{2\pi}|J,\theta;J',
\theta';K,\delta;l\rangle \langle J,\theta;J',\theta';K,\delta;l| d\eta(J,\theta)\cr
\cr
&&\times d\eta(J',\theta')d\mu(K,\delta)\cr
&&= \int_{0}^{\infty}\int_{0}^{\infty} \frac{J'^{l}}{ \rho_{1}(l)}\kappa^{n}e^{-J'} \sum_{n,m = 0}^{\infty}\frac{J^{\frac{n+m}{2}}}{\sqrt{\rho_{1}(n)\rho_{1}(m)}}\kappa^{n}e^{-J}
 \int_{0}^{2\pi}\int_{0}^{2\pi}e^{-i (\mathcal E_{n} -  \mathcal E_{m} )\theta}\frac{d\theta}{2\pi}
\frac{d\theta'}{2\pi}\crcr
&& \times dJdJ'|\Psi_{nl}\rangle \langle \Psi_{nl}|
\otimes \int_{0}^{L}
\sum_{k,q = 0}^{\infty}\frac{K^{\frac{k+q}{2}}}{\sqrt{\bar \rho(k) \bar\rho(q)}}
\varpi(K)dK \int_{0}^{2\pi}e^{i (\mathcal E'_{q} -  \mathcal E'_{k} )\delta}
\frac{d\delta}{2\pi}dK |\epsilon_{k}\rangle \langle \epsilon_{q}| \cr
&&=\sum_{n=0}^{\infty} \int_{0}^{\infty}\frac{J'^{l}}{l !} e^{-J'} dJ'
\int_{0}^{\infty}\frac{J^n}{n !}e^{-J}dJ|\Psi_{nl}\rangle \langle \Psi_{nl}| \otimes \sum_{k=0}^{\infty}\int_{0}^{L} \frac{K^{k}}{\bar \rho(K)}\varpi(K)dK
|\epsilon_{k}\rangle \langle \epsilon_{k}|.\nonumber
\\
\eeq
Using the definition of  the Gamma function, we obtain, on the Hilbert space  $\mathfrak H^{l}_{D} \otimes \tilde{\mathfrak H}$,  the following resolution of the identity:  
\beq
&&\int_{0}^{\infty}\int_{0}^{\infty}\int_{0}^{L}
\int_{0}^{2\pi}\int_{0}^{2\pi}\int_{0}^{2\pi}|J,\theta;J',
\theta';K,\delta;l\rangle \langle J,\theta;J',\theta';K,\delta;l| \cr
&& \times d\eta(J,\theta)d\eta(J',\theta')d\mu(K,\delta)
=\sum_{n=0}^{\infty}|\Psi_{nl}\rangle \langle \Psi_{nl}| \otimes
\sum_{k=0}^{\infty}|\epsilon_{k}\rangle \langle \epsilon_{k}| 
=  \1_{\mathfrak H^{l}_{D}} \otimes \1_{\tilde{\mathfrak H}},
\eeq
providing  that the relation (\ref{es28}) is satisfied.
$\hfill{\square}$

{\bf Proof of Proposition \ref{prop4.8}}
 From (\ref{es27}), and the measures  given by $d\eta(J',\theta') = d\nu(J')d\mu(\theta')$ as in (\ref{mes3}),
and  $d\sigma(J,\alpha_{k},\theta) = \varrho_{1}(J,\alpha_{k})dJd\mu(\theta), \; d\mu(K,\delta) = \varrho_{2}(K)dKd\mu(\delta), $
we have
\beq
&&\int_{0}^{\infty}\int_{0}^{\infty}\int_{0}^{L}
\int_{0}^{2\pi}\int_{0}^{2\pi}\int_{0}^{2\pi}|J,\theta;J',
\theta';K,\delta;l\rangle \langle J,\theta;J',\theta';K,\delta;l| d\sigma(J,\alpha_{k},\theta)\cr
\cr
&&\times d\eta(J',\theta')d\mu(K,\delta)
\cr
&&= \int_{0}^{\infty}\int_{0}^{\infty} \frac{J'^{l}}{ \rho_{1}(l)}\kappa^{n}
e^{-J'}
\sum_{n,m = 0}^{\infty}\frac{J^{\frac{n+m}{2}}}{\sqrt{\rho(n,\alpha_{k}) \rho(m,\alpha_{k})}} \frac{\varrho_{1}(J,\alpha_{k})}{\mathcal N(J,K)}
\int_{0}^{2\pi}\int_{0}^{2\pi}e^{-i (\mathcal E'_{n,\alpha_{k}} -  \mathcal E'_{m,\alpha_{k}} )\theta} \crcr
&&\times\frac{d\theta}{2\pi}\frac{d\theta'}{2\pi}dJdJ'
|\Psi_{nl}\rangle \langle \Psi_{nl}| \otimes \int_{0}^{L}
\sum_{k,q = 0}^{\infty}\frac{K^{\frac{k+q}{2}}}{\sqrt{\bar \rho(k) \bar \rho(q)}} \frac{\varrho_{2}(K)}{\mathcal N(K,J)}
\int_{0}^{2\pi}e^{i (\mathcal E'_{q} -  \mathcal E'_{k} )\delta}
\frac{d\delta}{2\pi}dK |\epsilon_{k}\rangle \langle \epsilon_{q}|\cr
\cr
&&=\sum_{n=0}^{\infty}\int_{0}^{\infty}\frac{J'^{l}}{l !}e^{-J'} dJ'
\int_{0}^{\infty}\frac{J^{n}}{\rho_{1}(n,\alpha_{k})\mathcal N(J,\alpha_{k})}  \varrho_{1}(J,\alpha_{k})dJ|\Psi_{nl}\rangle \langle \Psi_{nl}| \crcr
&&
 \otimes \int_{0}^{L}\sum_{k=0}^{\infty}  \frac{K^{k}}{\bar \rho(k)\mathcal N(K,J)}\varrho_{2}(K)dK|\epsilon_{k}\rangle \langle \epsilon_{k}|.
\eeq
Thereby,
\beq{\label{es32}}
&&
\int_{0}^{\infty}\int_{0}^{\infty}\int_{0}^{L}
\int_{0}^{2\pi}\int_{0}^{2\pi}\int_{0}^{2\pi}|J,\theta;J',
\theta';K,\delta;l\rangle \langle J,\theta;J',\theta';K,\delta;l| \cr
&&\times d\sigma(J,\alpha_{k},\theta)d\eta(J',\theta')d\mu(K,\delta)
=\sum_{n=0}^{\infty}|\Psi_{nl}\rangle \langle \Psi_{nl}| \otimes \sum_{k=0}^{\infty}
|\epsilon_{k}\rangle \langle \epsilon_{k}| = \1_{\mathfrak H^{l}_{D}} \otimes \1_{\tilde{\mathfrak H}},
\eeq
if there exist the densities $\varrho_{1}(J,\alpha_{k})$ and $\varrho_{2}(K)$ such that
\beq{\label{es34}}
\int_{0}^{\infty}\frac{J^{n}}{\rho_{1}(n,\alpha_{k})\mathcal N(J,\alpha_{k})}\varrho_{1}(J,\alpha_{k})dJ\int_{0}^{L} \frac{K^{k}}{\bar \rho(k)\mathcal N(K,J)}\varrho_{2}(K)dK  = 1,
\eeq
where
\beq
\varrho_{1}(J,\alpha_{k}) = \mathcal N(J,\alpha_{k})\varpi_{1}(J,\alpha_{k}), \; \varrho_{2}(K) = \mathcal N(K,J)\varpi_{2}(K)
\eeq
supplying that the measures $\varpi_{1}(J,\alpha_{k})$ and $\varpi_{2}(K)$ satisfy 
\beq
\int_{0}^{\infty}\frac{J^{n}}{\rho_{1}(n,\alpha_{k})}\varpi_{1}(J,\alpha_{k})dJ
\int_{0}^{L} \frac{K^{k}}{\bar \rho(k)}\varpi_{2}(K)dK  = 1.
\eeq
$\hfill{\square}$

{\bf Proof of Proposition \ref{prop4.10}} From the definition of the coherent states (\ref{el19}), we get
\beq
|J,\theta;J',\theta';\alpha_{k}\rangle \langle J,\theta;J',\theta';\alpha_{k}| 
&=& \frac{1}{\mathcal N(J) \mathcal N(J')}\sum_{n,l=0}^{\infty}\sum_{q,m=0}^{\infty}\frac{J^{\frac{n+q}{2}}J'^{\frac{l+m}{2}}
e^{i \left\{(q-n)\theta + (l-m)\theta'\right\}}}{\sqrt{n !l !q !m !}} \cr\cr
&& |\Psi_{nl}\rangle  \langle \Psi_{qm}| \otimes |\alpha_{k}\rangle \langle \alpha_{k}|,
\eeq
such that
\beq
&&\int_{0}^{2\pi}\int_{0}^{2\pi}|J,\theta;J',\theta';\alpha_{k}\rangle \langle J,\theta;J',\theta';\alpha_{k}|
d\mu(\theta)d\mu(\theta')\mathcal N(J)\mathcal N(J') \cr
\cr
&& = \sum_{n,l=0}^{\infty}\sum_{q,m=0}^{\infty}\frac{1}{\sqrt{n !l !q !m !}}\delta_{qn}\delta_{lm}J^{\frac{n+q}{2}}J'^{\frac{l+m}{2}} |\Psi_{nl}\rangle  \langle \Psi_{qm}| \otimes |\alpha_{k}\rangle \langle \alpha_{k}| \cr
&& = \sum_{n,l=0}^{\infty}\frac{1}{n !l !}J^{n}J'^{l}|\Psi_{nl}\rangle
\langle \Psi_{nl}| \otimes |\alpha_{k}\rangle \langle \alpha_{k}|.
\eeq
Thereby
\beq
&&
\int_{0}^{\infty}\int_{0}^{\infty}\int_{0}^{2\pi}\int_{0}^{2\pi}
|J,\theta;J',\theta';\alpha_{k}\rangle \langle J,\theta;J',\theta';\alpha_{k}|
d\mu(\theta)d\mu(\theta')\mathcal N(J)\mathcal N(J') \cr
\cr
&& \times d\nu(J)d\nu(J')\cr
&& = \sum_{n,l=0}^{\infty}|\Psi_{nl}\rangle  \langle \Psi_{nl}|
\otimes |\alpha_{k}\rangle \langle \alpha_{k}|
= \1_{\mathfrak H_{D}} \otimes \1_{\tilde{\mathfrak H}_k}.
\eeq
$\hfill{\square}$

{\bf Proof of Proposition \ref{prop4.11}}
Let us set  
\beq 
H_{1_{OSC}} - H_{2_{OSC}} = \sum_{n,l=0}^{\infty} \omega_{c}(n-l)
|\Psi_{nl};\alpha_{k}\rangle \langle \Psi_{nl};\alpha_{k}|, 
\eeq
with $|\Psi_{nl};\alpha_{k}\rangle  :=
|\Psi_{nl}\rangle
\otimes |\alpha_{k}\rangle$. Then, we have
\beq
H|J,\theta;J',\theta';\alpha_{k}\rangle  &=& \frac{1}{\left[\mathcal N(J) \mathcal N'(J)\right]^{1/2}}\sum_{n,l=0}^{\infty}
\sum_{q,m = 0}^{\infty}\frac{\omega_{c} (n-l) J^{q/2}J'^{m/2}e^{-i (q\theta - m\theta')}}{\sqrt{q !m !}} \cr
&&  \times \langle \Psi_{nl}|\Psi_{qm}
\rangle|\Psi_{nl}\rangle \otimes \langle \alpha_{k}|\alpha_{k}\rangle |\alpha_{k}\rangle \cr
                                    &=& \frac{1}{\left[\mathcal N(J) \mathcal
N'(J)\right]^{1/2}}\sum_{n,l=0}^{\infty}\frac{\omega_{c} (n-l) J^{n/2}J'^{l/2}e^{-i (n\theta - l\theta')}}{\sqrt{n !l !}}
|\Psi_{nl}\rangle \otimes |\alpha_{k}\rangle.
\eeq
Thus,
\beq
&&\langle J,\theta;J',\theta';\alpha_{k}| H |J,\theta;J',\theta';\alpha_{k}\rangle \cr
&& = \frac{1}{\left[\mathcal N(J) \mathcal N'(J)\right]}
\sum_{n,l=0}^{\infty}
\sum_{q,m = 0}^{\infty}\frac{\omega_{c} (n-l) J^{\frac{n+q}{2}}J'^{\frac{l+m}{2}}
e^{-i ((q-n)\theta - (l-m)\theta')}}
{\sqrt{n! l !q !m !}} \delta_{qn} \delta_{lm} \cr
&& = \omega_{c} \left[ J \frac{1}{\mathcal N(J)}\sum_{n=0}^{\infty}\frac{J^{n}}{n !} - J' \frac{1}{\mathcal N(J')}\sum_{l=0}^{\infty}\frac{J'^{l}}{l !}\right] = \omega_{c} (J - J').
\eeq
$\hfill{\square}$

\end{document}